\documentclass[reprint,superscriptaddress,showpacs,amsmath,amssymb,aps,pre]{revtex4-1}
\usepackage{graphicx}
\usepackage[justification=raggedright]{caption} 

\usepackage{epstopdf}
\usepackage{amsmath}
\usepackage{diagbox} 
\usepackage{booktabs}
\usepackage{array}         
\usepackage{multirow}     
\usepackage{makecell}     
\usepackage[colorlinks=true, citecolor=blue, urlcolor=blue, linkcolor=blue]{hyperref}
\usepackage{subcaption}
\usepackage{algpseudocode}
\usepackage{xcolor}
\usepackage{algorithmicx}
\usepackage{algpseudocode}
\usepackage{amsmath}
\usepackage[ruled,vlined]{algorithm2e}

\begin{document}
\title{Decoding species coexistence: A reinforcement learning perspective}

\author{Kaiwen Jiang}
\affiliation{School of Physics and Information Technology, Shaanxi Normal University, Xi'an 710061, P. R. China}
\author{Chenyang Zhao}
\affiliation{School of Physics and Information Technology, Shaanxi Normal University, Xi'an 710061, P. R. China}
\affiliation{School of Physics and Electronic Science, East China Normal University, Shanghai 200241, P. R. China}
\author{Shengfeng Deng}
\email[Email address: ]{dengsf@snnu.edu.cn}
\affiliation{School of Physics and Information Technology, Shaanxi Normal University, Xi'an 710061, P. R. China}
\author{Weiran Cai}
\affiliation{School of Computer Science, Soochow University, Suzhou 215006, P. R. China}
\author{Jiqiang Zhang}
\affiliation{School of Physics, Ningxia University, Yinchuan 750021, P. R. China}
\author{Li Chen}
\email[Email address: ]{chenl@snnu.edu.cn}
\affiliation{School of Physics and Information Technology, Shaanxi Normal University, Xi'an 710061, P. R. China}

\date{\today}

\begin{abstract}
A central goal in ecology is to understand how biodiversity is maintained. Previous theoretical works have employed the rock-paper-scissors (RPS) game as a toy model, demonstrating that population mobility is crucial in determining the species' coexistence. One key prediction is that biodiversity is jeopardized and eventually lost when mobility exceeds a certain value — a conclusion at odds with empirical observations of highly mobile species coexisting in nature. To address this discrepancy, we introduce a joint reinforcement learning framework to study a spatial RPS model, where individuals' mobility for each species is not fixed but is guided by a common experience pool in the form of a Q-table, and its members jointly revise it via a Q-learning algorithm. 
Our results show that all three species can coexist stably, with extinction probabilities remaining low across a broad range of baseline migration rates. Mechanistic analysis reveals that individuals develop two behavioral tendencies: survival-priority (escaping from predators) and predation-priority (remaining near prey). While species coexistence emerges from the balance of the two tendencies, their imbalance jeopardizes biodiversity. Notably, there is a symmetry-breaking of action preference in a particular state that is responsible for the divergent species densities. Furthermore, when Q-learning species interact with fixed-mobility counterparts, those with adaptive mobility exhibit a significant evolutionary advantage. 
Our study suggests that joint reinforcement learning offers a promising new perspective for uncovering the mechanisms of biodiversity and designing conservation strategies.
\end{abstract}

\maketitle
\section{introduction}\label{sec:introduction}

Ecological systems are crucial for humans, as they provide materials and energy for our survival~\cite{Assessment2005ecosystems}, and biodiversity is the key property that supports their functional working. Since Darwin first envisioned the ``tree of life"~\cite{Darwin2004origin}, understanding the mechanisms underpinning species coexistence has remained a central challenge in ecology~\cite{Pennisi2005determines}. According to the Dasgupta report~\cite{Dasgupta2021}, biodiversity is declining faster than at any time in human history; the extinction rate of species nowadays is around 100 to 1000 times higher than the baseline value. Decoding how species coexist is a grand scientific question that can help preserve biodiversity and ultimately promote the sustainability of human civilization.

The past several decades have witnessed significant progress in theoretical ecology~\cite{May2007theoretical} through the combination of nonlinear dynamics, agent-based modeling, and evolutionary game theory. While the classical population models, such as the Lotka-Volterra model~\cite{Murray2002mathematical}, can provide a deterministic description of the system in the form of differential equations and can be elegantly solved, they fail to capture the fluctuations and complex interactions~\cite{Lehman1997competition}. Agent-based models complement the macroscopical method by allowing for more details and revealing diverse spatiotemporal patterns~\cite{Mclane2011role}.

Evolutionary game theory~\cite{1982Evolution,Hofbauer1998,Nowak2006} contends that the success of one species intrinsically depends on the behavior of others, providing a powerful theoretical framework for population dynamics. Within this context, the rock-paper-scissors (RPS) game has emerged as a canonical model for species diversity \cite{Durrett1997,Durrett1998,kerr2002,Czaran2002,May1975,Johnson2002,Reichenbach2006,szabo2007,Szolnoki2014cyclic,Zhou2016rock}, where rock is wrapped by paper, paper is cut by scissors, and scissors are crushed by rock. This nonhierarchical, cyclic competition structure captured in RPS game is widely observed in nature, such as lizard populations~\cite{Sinervo1996rock}, strains of yeast~\cite{Paquin1983relative}, reef invertebrates~\cite{Jackson1975alleopathy}, among others~\cite {Czaran2002chemical}. 
Intuition suggests diversity should persist in such systems: an endless pursuit where each species dominates one competitor yet is dominated by another, creating a cyclic hierarchy of advantage.
 However, this is not necessarily the case when individuals are located in a spatial domain, where they move constantly.
 In the seminal work by Reichenbach et. al.~\cite{2007Mobility}, they incorporate mobility into a spatially extended RPS model (referred to as the RMF model afterwards), and reveal that the three species coexist in the form of spiral waves for low mobility, but extinction occurs when their mobilities exceed a critical value. This prediction is, however, inconsistent with reality, as there are many examples where species with high mobility in nature coexist well with each other~\cite{ritchie2002competition}.

Subsequent research proposes some mechanisms aiming to fill this gap by introducing new ingredients, such as intraspecific competition~\cite{2010Yang}, viral/infectious transmission~\cite{2010Wang}, cross-patch migration~\cite{2011Wang}, habitat suitability~\cite{2013Persistent}, among others~\cite{Huang2023Fitness,Lee2022Costly,Menezes2022Adaptive,Park2020Relativistic,Park2019Fitness}. In particular, in Ref.~\cite{2013Persistent} {Park} \emph{et al.} introduce an index to characterize the local habitat suitability whereby individuals adjust their migration; they find robust coexistence even in the high-mobility regime.  This work captures a basic biological instinct -- individuals are likely to move away when the local habitat becomes hostile and exhibit low mobility in favorable surroundings otherwise. 
While these works correctly grasp the adaptive nature in mobility, their models hinge on handcrafted heuristic rules that fail to capture the learning processes, which are inherent to most living organisms and whereby the adaptivity is acquired~\cite{Staddon1983adaptive}.

Recently, reinforcement learning (RL), a fundamentally different paradigm, has offered new perspectives on understanding both social and ecological systems. It has been shown that the emergence of cooperation~\cite{Ding2023Emergence,zheng2024evolution,xu2024reinforcement, Zhang2024emergence,xing2025online}, trust~\cite{Zheng2024decoding}, fairness~\cite{zheng2025decoding}, and resource allocation~\cite{zheng2025optimal}, and some other human behaviors~\cite{Zhang2020Oscillatory,Zhang2020Understanding} can be well understood with RL. Unlike mechanical models, where individuals make their moves according to some prescribed rates or probabilities, RL individuals score different actions within different states, which are determined by their environment. This allows them to adjust their actions adaptively and could make completely different moves in response to different surroundings. A key idea behind RL individuals is that they aim to maximize the accumulated payoffs rather than the immediate rewards, thereby better adapting to their surroundings. 

However, the studies applying RL in ecology mostly focused on the predator-prey systems~\cite{2015Co-evolution,2019Deep-Reinforcement,2020Wang,2021Co-Evolution}, with particular interests in swarming behaviors~\cite{2023Predator-prey} and collaborative hunting~\cite{2024Collaborative}. A recent work~\cite{si2025beyond} based on the prisoner's dilemma game examines a three-species population using Q-learning, with a focus on sustaining cooperation. 
{Another research~\cite{mangold2025dilution} studied dilution and mobility in a population with Q-learning powered spatial prisoner's dilemma, revealing their great impact and a symbiotic effect.} Importantly, a common assumption in most of these works is that learning is posed at the individual level, neglecting the fact that for individuals of many species they share the experiences from their ancestors, and jointly update their experience~\cite{falcon2019collective,kameda2022information}.
Therefore, \emph{can the RL paradigm offer new insights into biodiversity, and can the learning be performed collectively?} This allows the RL to occur at the species level, where individuals of the same species share and jointly revise a common experience pool, which in turn guides their mobilities. By doing so, this would provide a more natural explanation for the gap left by the RMF model and answer the question of how species with high mobility coexist.

In this work, we propose the RL paradigm to identify the mechanism of species coexistence. Specifically, we employ a joint Q-learning algorithm on a spatial RPS model; individuals belonging to the same species are guided by a common Q-table. This shared Q-table can be interpreted as the collective wisdom passed from their ancestors. For simplicity, individuals are engaged in random exploration to ensure the three Q-tables converge in the learning stage; in the later stage, the evolution of the three species is guided by their respective Q-tables. Surprisingly, we uncover that species empowered by RL coexist very well even in the high baseline mobility region, where extinction is certain in the RMF model. Preference analysis reveals that a balanced priority in escape and predation ruins the spiral waves and sustains their coexistence. Further studies of mixed populations reveal the obvious advantage of Q-learning species over traditional species (with fixed mobility) and the rich dynamics in a heterogeneous Q-learning population with diverse preferences.

The rest of the paper is organized as follows: 
Sec.~\ref{sec:model} presents our spatial RPS model implemented with a joint Q-learning algorithm. 
Sec.~\ref{sec:result} shows the evolutionary outcomes of the three species along with the results for the traditional RMF for comparison. 
Sec.~\ref{sec:mechanism} provides the mechanism analysis explaining species coexistence. 
Sec.~\ref{sec:extension} presents two model extensions, one for the mixture of traditional species with our Q-learning species, the other for all Q-learning species but with diverse preferences. 
Sec.~\ref{sec:conclusion} concludes this study.

\begin{figure}
\centering
\includegraphics[width=0.5\linewidth]{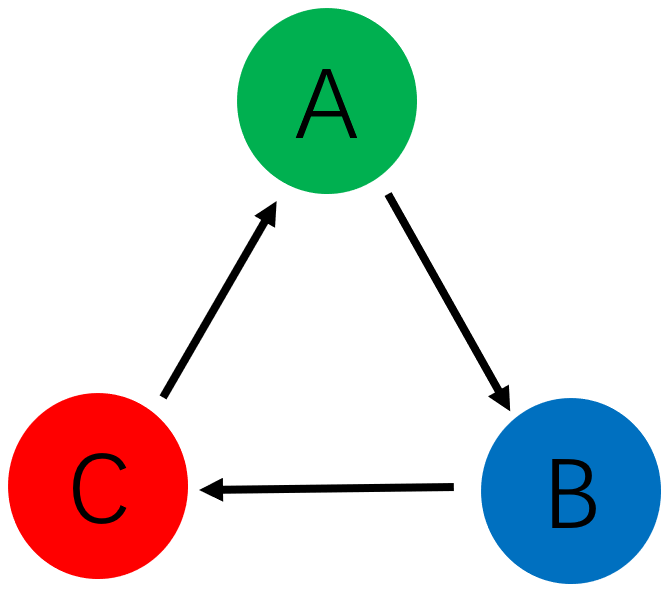}
\caption{\textbf{Rock-paper-scissors game.} Three species, A, B, and C, are cyclically dominant over each other.}
\label{fig:RPS}
\end{figure}

\section{model} \label{sec:model}

We start with the standard cyclic competition structure among three species, denoted as A, B, and C, which are governed by the rock-paper-scissors (RPS) game, shown in Fig.~\ref{fig:RPS}. Specifically, we consider a spatial version of RPS dynamics~\cite{2007Mobility}, where the individuals of the three species occupy the intersection points of a square lattice of size $N=L\times L$ with periodic boundary conditions. There, the spatial evolution contains interspecific competition, reproduction, and migration, which can be summarized by the following reactions:
\begin{equation}\label{eq:1}
A B \xrightarrow{\sigma} A \varnothing, \quad B C \xrightarrow{\sigma} B \varnothing, \quad C A \xrightarrow{\sigma} C \varnothing,
\end{equation}
\begin{equation}\label{eq:2}
A \varnothing \xrightarrow{\mu} A A, \quad B \varnothing \xrightarrow{\mu} B B, \quad C \varnothing \xrightarrow{\mu} C C,
\end{equation}
\begin{equation}\label{eq:3}
A \Box  \xrightarrow{\varepsilon  } \Box  A, \quad B \Box  \xrightarrow{\varepsilon} \Box  B, \quad C \Box  \xrightarrow{\varepsilon} \Box  C.
\end{equation}
Here $\varnothing$ denotes an empty site and $\Box$ represents any species or an empty site.
Reactions~(\ref{eq:1}) describe cyclic predation processes between different species, as shown in Fig.~\ref{fig:RPS}, which occur at a rate $\sigma$. 
Reactions~(\ref{eq:2}) show the reproduction process with a rate $\mu$, which can only take place when an adjacent site is empty. 
Reactions~(\ref{eq:3}) represent the migration process with an exchange rate $\varepsilon _{0}$. 
Following these reactions, one can monitor the density evolution of the three species and study the impact of these parameters, e.g., with the Gillespie algorithm~\cite{2007Mobility}. Typically, the predation and reproduction rates (i.e., $\sigma$ and $\mu$) are assumed to be fixed, and previous works have extensively studied the impact of migration~\cite{2007Mobility,Reichenbach2008self}.

\begin{table}
	\centering
	\caption[Q-table]{\textbf{Q-table for each species}. The state is jointly defined by the number of prey $n_{\text{prey}}$ and predators $n_{\text{predator}}$ in its four nearest neighboring sites, i.e., $s = \left( n_{\text{prey}}, n_{\text{predator}} \right)$. Actions consist of seven migration willingness $\lambda\in\mathcal{A}=\{-3,-2,-1,0,1,2,3\}$, {corresponding to seven different migration rates.}}
	\setlength{\heavyrulewidth}{1.5pt} 
	\setlength{\lightrulewidth}{1pt}   
	\begin{tabular}{>{\bfseries}c|>{\bfseries}c>{\bfseries}c>{\bfseries}c>{\bfseries}c}
		\toprule
		\diagbox [width=5em,trim=l] {State}{Action} & $\lambda=-3\left ( a_{1} \right )$ & $\lambda=-2\left ( a_{2} \right )$ & $\dots$ & $\lambda=3\left ( a_{7} \right )$  \\
		\hline
		$s_{1}=\left ( 0,0 \right )$ & $Q_{s_{1},a_{1}}$& $Q_{s_{1},a_{2}}$& $\dots$ & $Q_{s_{1},a_{7}}$  \\
		$s_{2}=\left ( 0,1 \right )$ & $Q_{s_{2},a_{1}}$& $Q_{s_{2},a_{2}}$& $\dots$ &$Q_{s_{2},a_{7}}$  \\
		$\vdots$ & $\vdots$ & $\vdots$ &$\ddots$  & $\vdots$ \\
		$s_{15}=\left ( 4,0 \right )$ & $Q_{s_{15},a_{1}}$ & $Q_{s_{15},a_{2}}$ & $\dots$ & $Q_{s_{15},a_{7}}$  \\
		\bottomrule
	\end{tabular}\vspace{0cm}
	\label{table:1}
\end{table}

Here, we instead resort to Q-learning~\cite{Q-learning1,Q-learning2}, a classic reinforcement learning algorithm.
Unlike most previous practices, where each individual is associated with a Q-table, in our study, individuals belonging to a species share a common Q-table that guides their migration. The Q-table can be interpreted as the collective wisdom for a given species that jointly emerges and guides its members' decision-making. {The Q-table is a two-dimensional table expanded by the state set $\mathcal{S}$ and the action set $\mathcal{A}\!=\!\{-3,-2,-1,0,1,2,3\}$, shown in Table~\ref{table:1}, where the discrete migration willingness is introduced to adjust the migation rate}. The state serves to depict the local environment, defined by the number of prey and predators in the four neighbors around the focal individual; i.e., $s=(n_\text{prey},n_\text{predator})$, so the state set  $\mathcal{S}=\{ s_{1}=(0,0), s_{2}=(0,1),\dots,s_{15}=(4,0)\}$ captures all possible neighborhood regarding predation. The actions of the individual are comprised of a series of migration willingness $\lambda$, {and the migration rate is accordingly defined as}
\begin{equation}\label{eq:4}
\varepsilon= \varepsilon_{0}\exp(\beta \lambda),
\end{equation}
where $\lambda\in\mathcal{A}$, $\beta$ is a temperature-like parameter and is fixed at $\beta=2$ in our study. 
{Although migration rates are continuous in reality, here we discretize the action space as required by Q-learning, and $\beta$ controls the degree of their separation.}
It can be seen that when $\lambda > 0$, individuals move faster than the benchmark migration rate $\varepsilon_{0}$, and become slower in the opposite case, $\lambda < 0$. Unlike previous studies, where the migration rate $\varepsilon$ is uniform for all individuals (i.e., $\varepsilon=\varepsilon_{0}$), this value could vary among individuals and is time-evolving. In our practice, the migration rate for two neighboring sites, say site $i$ and $j$, the rate for their position exchange is set to be $\varepsilon_{ij}=\varepsilon_{0}\exp (\beta \lambda_{ij})$, where $\lambda_{ij}=(\lambda_{i}+ \lambda_{j})/2$ if site $j$ is occupied and $\lambda_{ij}=\lambda_{i}$ if $j$ is empty~\cite{2013Persistent}.

Importantly, the items $Q_{s,a}$ in the table are termed the action-value function, estimating the value of the action $a$ within the given state $s$, which can be taken as a measure of action preference. The larger the value of $Q_{s,a}$, the stronger the preference in action $a$ for the individual within state $s$. By scoring different actions within different states, Q-tables guide individuals to migrate properly. 
Compared to the Q-learning at the individual level, our joint Q-learning for species could be more reasonable because most species' behaviors are guided by common wisdom from their groups that is accumulated for many generations, rather than learn everything \emph{ab initio} for individuals~\cite{Staddon1983adaptive}.

\begin{figure*}[htbp]
\centering
\includegraphics[width=\linewidth]{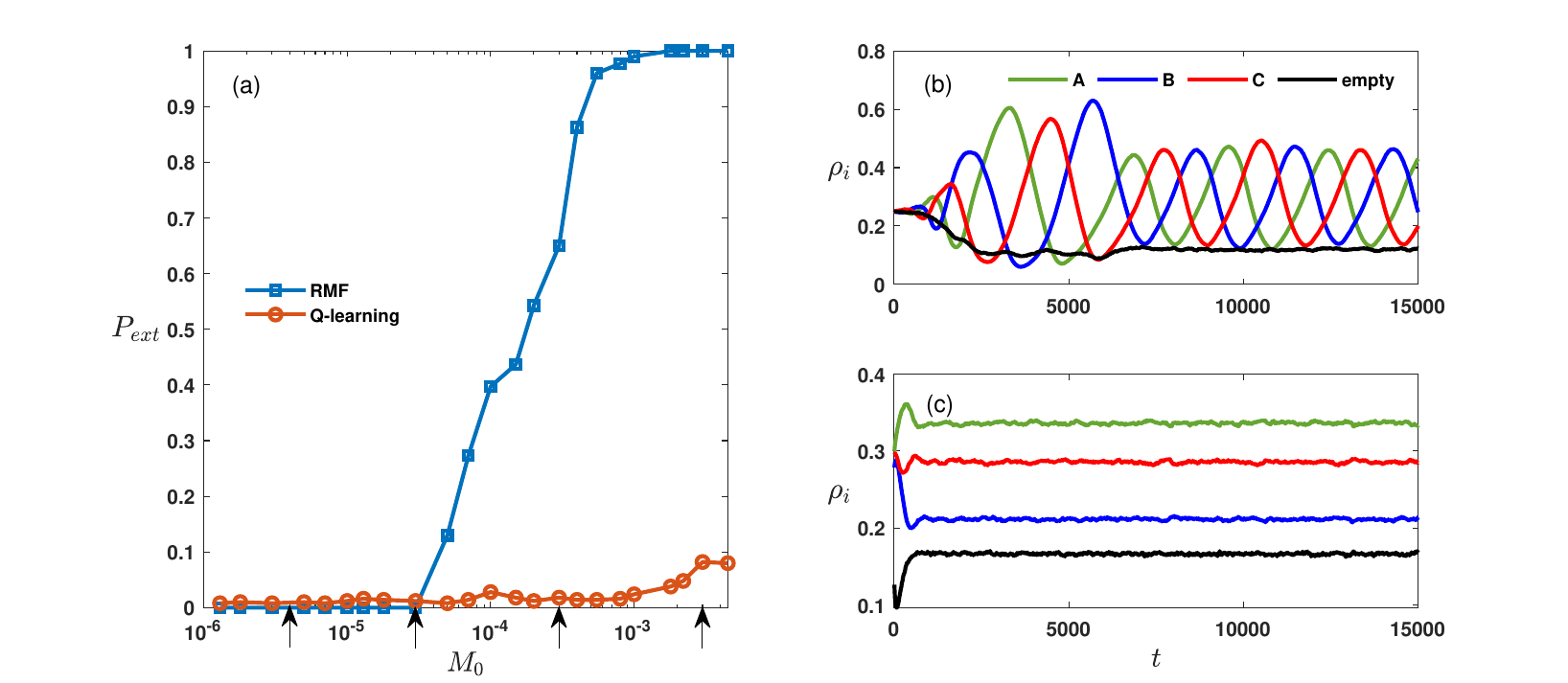}
\caption{\textbf{Extinction probability and typical time series of species densities.} 
(a) Extinction probability versus the baseline mobility $M_{0}$. The blue squares and red circles represent the results for the traditional RMF model and our Q-learning model, respectively. The extinction probability drops significantly after applying the Q-learning algorithm, greatly enhancing system stability. Each data point is computed by averaging over 1000 realizations with an evolution duration of $2N$.
Typical time series of three species densities as well as the density of empty sites for the traditional RMF model (b) and our model (c), both at $M_{0}=3\times10^{-4}$. This means that when species are empowered with reinforcement learning, they are much better at coexisting with each other than the previous RMF model, and the density oscillation is suppressed.
Parameters: ($R_{p}=2$ and $R_{s}=0.5$) , and $N=100\times100$.
}
\label{fig:Pext}
\end{figure*}

Without loss of generality, each site of the lattice at the beginning is randomly occupied by an individual of type A, B, C, or left empty,  meaning a finite carrying capacity. The initial action for each is randomly chosen $a_i\in\mathcal{A}$, and the elements $Q_{s,a}$ for the three Q-tables are randomly initialized to a value between 0 and 1 independently. 
The evolution follows a synchronous updating procedure. At round $t$, there is a reaction among all possible Reactions (\ref{eq:1}-\ref{eq:3}) going to occur with probability proportional to their rates. The reward for a successful predation for the predator is $R_p$ (i.e., Reactions (\ref{eq:1})), and the reward for survival for one round is denoted as $R_s$.
Different from the classic Q-learning, where the gaming and learning processes are repeated iteratively, here the two processes are conducted separately for simplicity. The learning process unfolds in the first stage until the three Q-tables are converged; then, the population's migration strictly follows the guidance of Q-tables for the second stage of evolution.

Specifically, in the learning process, every individual makes a random action $a\in\mathcal{A}$ for migration, and their experiences are accumulated by updating the Q-table belonging to their species as follows: 
\begin{equation}
\label{eq:Bellman_eq}
\begin{split}
&Q_{s,a}(t+1) =  \frac{1}{|\mathcal{N}_{m}|} \sum_{j\in\mathcal{N}_{m}} \{ Q_{s,a}(j)+\alpha [ R(j) \\
                    & + \gamma \max_{a^{'} } Q_{s^{'},a^{'}}(j) - Q_{s,a}(j)]\},
\end{split}
\end{equation}
where $s$ and $a$ represent the current state and the action that the focal individual has just taken, and $s^{'}$ is the new state in round $t+1$. 
The parameter $\alpha \in \left ( 0,1 \right ]$ is the learning rate, which determines the contribution to Q-value from the current round. $\gamma \in \left [ 0,1 \right )$ is the discount factor, which captures the weight of future rewards, where $\max_{a^{'} } Q_{s^{'},a^{'}}$ denotes the expected maximal Q-value in round $t+1$ within $s^{'}$.
$R$ is the total reward, including the reward for a successful predation $R_{p}$ and the survival $R_{s}$. 
 $\mathcal{N}_{m}\in\{ \mathcal{N}_{A},\mathcal{N}_{B}, \mathcal{N}_{C}\}$, denotes the individual set of the three species, 
 {and $|\mathcal{N}_{m}|$ is the the population size of the corresponding species.}
 Eq.~(\ref{eq:Bellman_eq}) shows that the Q-table for a given species $\mathcal{N}_{m}$ is jointly revised in an average manner by all individuals that belong to it. 
 Importantly, this learning scheme adopts the $\epsilon$-greedy Q-learning with $\epsilon=1$; its advantage is that the Q-tables can be rapidly converged, as different states can be visited more frequently than in the conventional setup with a small $\epsilon$. 
After the three Q-tables converge, the evolution enters the second stage, where their respective Q-tables strictly guide the migration of the population, and the three Q-tables are no longer revised.
The evolution of the population is terminated when it reaches equilibrium or the desired duration.

In our practice, a transient of 5000 steps is used for each realization, and both Q-tables and the migration rate are updated every 10 steps to guarantee that substantial changes have happened in the neighborhood before the Q-table revision or migration adjustment.
For more details, we provide more description and the pseudocode (Algorithm~\ref{algorithm:1}) in Appendix~\ref{Appendix:pseudocode}.   
As an ecological system, it's natural to monitor the densities of the three species, denoted as $\rho_{A, B, C}$. Extinction occurs when at least one of the three densities becomes zero. Since the evolution is stochastic, we compute the frequency of extinction as the extinction probability $P_{ext}$.
To be consistent with previous studies~\cite{2007Mobility,Reichenbach2008self}, we also adopt the macroscopic diffusion constant $M_{0} =\varepsilon _{0} (2N ) ^{-1}$ as the control parameter for mobility.

\section{results}\label{sec:result}

We first report the dependence of extinction probability $P_{ext}$ on the mobility $M_0$ with the system size of $N=100\times100$, and compare the result to the case in RMF model~\cite{2007Mobility} where the migration is constant, shown in Fig. \ref{fig:Pext}.
Fig.~\ref{fig:Pext}(a) shows that the extinction probability remains low $P_{ext}<10\%$ for the whole studied range, and only a slight increase in $P_{ext}$ is detected at the high mobility end. 
This is in sharp contrast with the observations in the RMF model, where the extinction probability transition occurs at around $M_0^c \approx 4.5 \times 10^{-4}$ and is significantly increased when $M_0$ becomes larger, $P_{ext}\rightarrow 1$ when $M_0\gtrsim10^{-3}$. This means that when individuals adopt Q-learning to make the decision of migration, the extinction is significantly reduced, and thus, the biodiversity is properly preserved.

Fig.~\ref{fig:Pext}(b, c) provide the typical time series for the two scenarios at $M_0=3\times10^{-4}$ of all densities $\rho_i$, where $i\in\{A, B, C, \varnothing\}$. Fig.~\ref{fig:Pext}(b) shows that the densities in the RMF model are strongly oscillatory; they wane and wax all the time after the transient. By contrast, the densities in our Q-learning model present only slight fluctuations after the transient, as shown in Fig.~\ref{fig:Pext}(c). As expected, too strong an oscillation in species density $\rho_i$ leads to extinction, and the reduced oscillation promotes species coexistence. 
Though the densities for the three species are not identical, meaning there is an underlying symmetry-breaking in the evolution, which will be explained in Sec.~\ref {sec:symmetry-breaking}. 

\begin{figure*}[htbp]
\centering
\includegraphics[width=\textwidth]{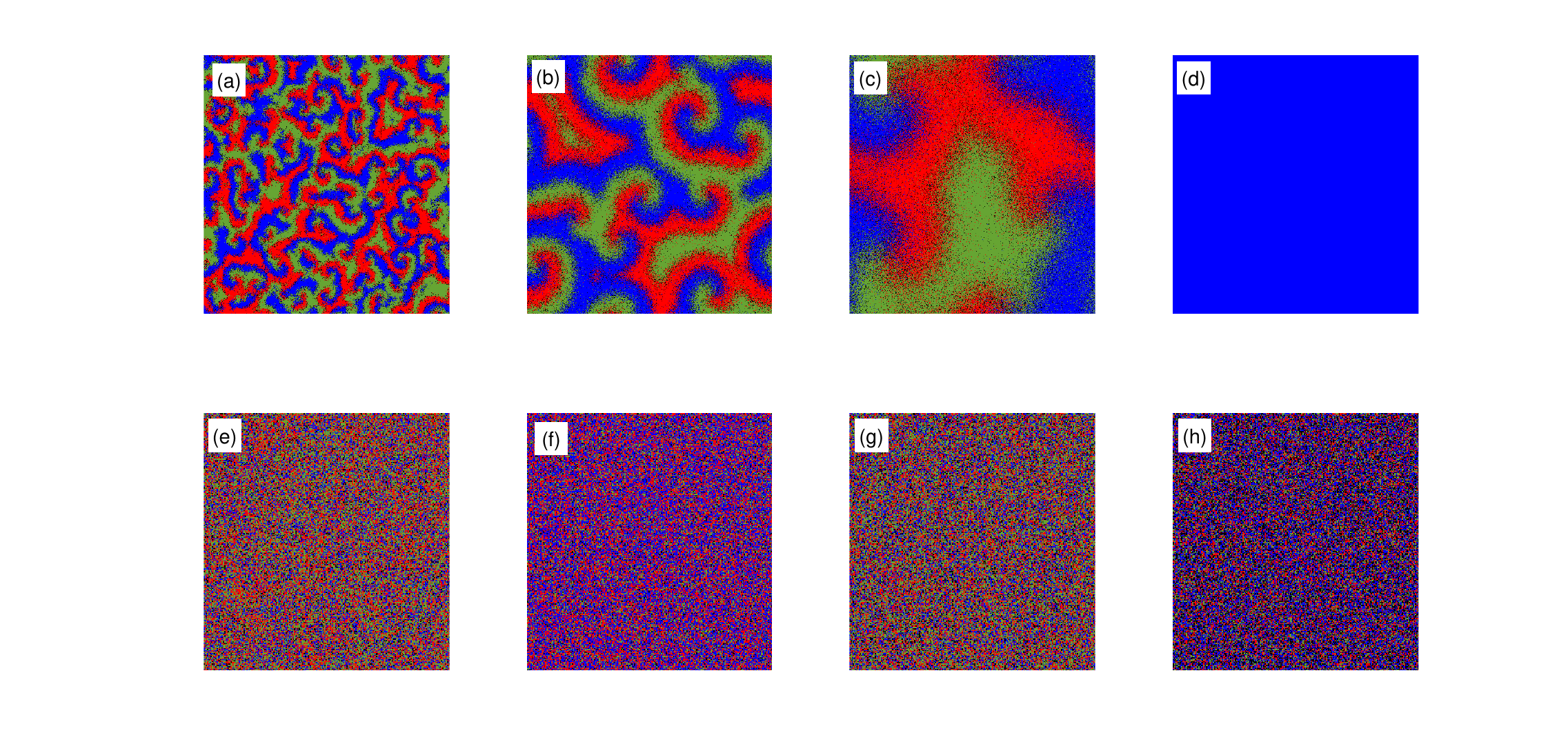}
\caption{\textbf{Typical spatial patterns.} 
The first row shows the patterns for the RMF model, and the second row is for our Q-learning RPS model. Sites with species A, B, C, and empty are represented by green, blue, red, and black, respectively.
Four columns correspond to four mobility values: (a, e) $M_{0}=3\times10^{-6}$, (b, f) $M_{0}=3\times10^{-5}$, (c, g) $M_{0}=3\times10^{-4}$, and (d, h) $M_{0}=3\times10^{-3}$, as indicated by the arrows in Fig.~\ref{fig:Pext}(a). 
Comparison reveals that the spiral waves in the RMF model disappear when the individuals are empowered with Q-learning.
Parameters: $N=500\times500$, and the snapshots are sampled at the end of $2N$ evolutionary steps. 
}	
\label{fig:pattern}
\end{figure*}

To develop some intuition for this distinction, some spatial snapshots are illustrated in Fig.~\ref{fig:pattern}; the top row is for the RMF model and the bottom row for our Q-learning model, respectively. The four columns correspond to four baseline migration rates: $M_{0}=3\times10^{-6}, 3\times10^{-5}, 3\times10^{-4}$, and $3\times10^{-3}$ for both scenarios, indicated by the four arrows in Fig.~\ref{fig:Pext}(a). As can be seen, spiral waves are emerging for the constant migration scenario, and the characteristic size of these waves increases with the migration rate $M_0$. As the characteristic size increases to be close to the domain dimensions, these species clusters likely become extinct due to the finite-size effect. An extinction example is shown in Fig.~\ref{fig:pattern}(d), where one species initially disappears, and the prey individuals of the remaining two species are eventually consumed up by the predator population, as they have no predators left. This picture is well-established in previous studies~\cite{2007Mobility,Reichenbach2008self}.

 In contrast, for the Q-learning RPS model, no spiral waves are formed; instead, individuals of different types are evenly dispersed throughout the domain. This is still the case even at high migration rates where $M_0 > 10^{-3}$ (e.g., Fig.~\ref{fig:pattern}(h)), and all three species still coexist.
These patterns starkly contrast with those in the RMF model, meaning that the adaptive adjustment of migration by Q-learning can avoid the formation of spiral waves, which in turn promotes the species coexistence, even with a large baseline migration rate, where biodiversity is impossible in the traditional RMF model.

\begin{figure*}
\centering
\includegraphics[width=0.4\linewidth]{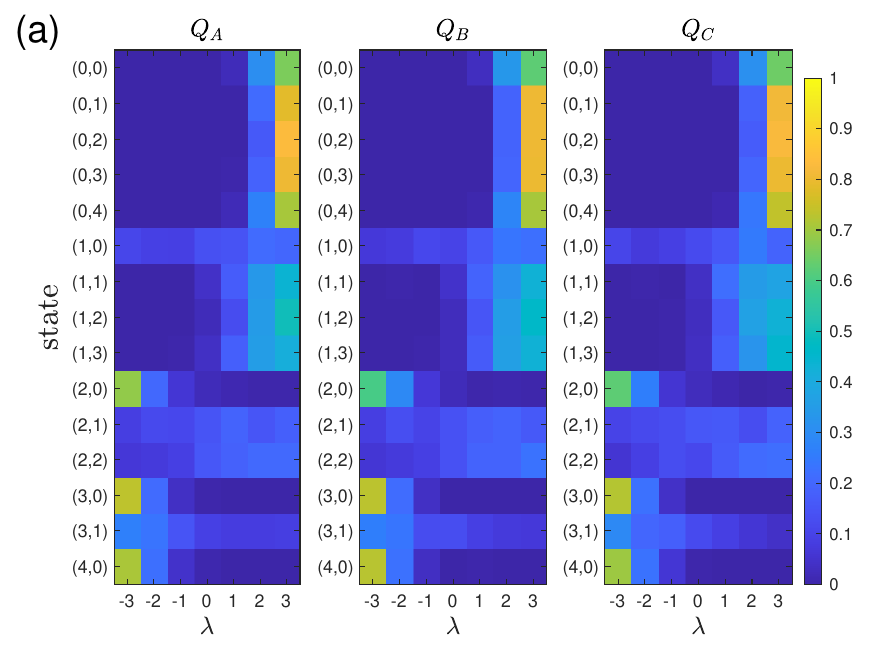}
\includegraphics[width=0.4\linewidth]{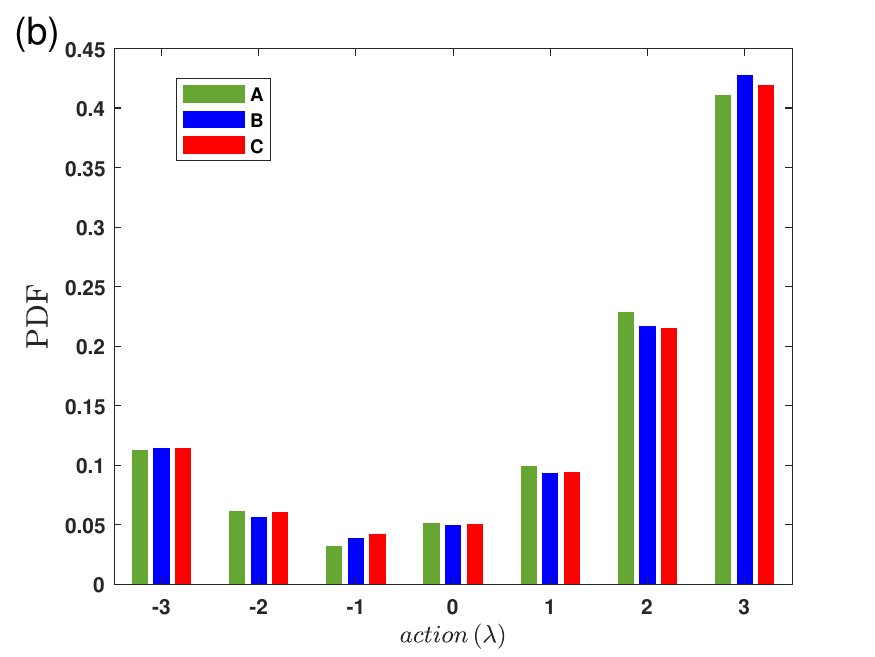}
\caption{\textbf{Distribution of action preferences.} 
(a) Color-coded action preferences for the three species. Three subplots correspond to the probabilities of action being chosen within all states for species A, B, and C, respectively. 
{The preferences are respresented as probabilities, each corresponding to one of $15\times 7$ state-action combinations.} 
A total of 1,000 independent runs are performed, and the statistics are computed at the end of the learning stage. These probabilities are normalized for each state $s\in\mathcal{S}$.  
(b) Probability density function (PDF) of different actions adopted $\lambda$ for the three species in the stable state. 
It can be observed that their distributions exhibit bimodality, where the large and small migrations are more preferred. 
Together, these plots reveal two tendencies: survival priority -- moving away from predators, and predation priority -- staying near prey.
Parameters: $R_{p}=2$, $R_{s}=0.5$, $M_{0}=5\times10^{-4}$ and $N=300\times300$.
}
\label{fig:balanced}
\end{figure*}

\section{mechanism analysis}\label{sec:mechanism}

To understand how joint Q-learning promotes biodiversity,  we turn to mechanism analysis by examining the action preference of the three species.
Fig.~\ref{fig:balanced}(a) computes the probabilities for every action $\lambda\in\mathcal{A}$ being chosen within each state $s\in\mathcal{S}$, which captures the action preference of the three species through Q-learning. These probability distributions reveal a symmetric structure in the action preference, where all three species exhibit similar patterns of action preference.
These distributions demonstrate two prominent tendencies: (i) individuals prefer to escape when predators are in their neighborhood, which we term ``survival-priority"; (ii) and to stay put when prey are around, which we term ``predation-priority". These two tendencies are most prominent for the two ends at $\lambda=\pm3$, see Fig.~\ref{fig:balanced}(a).

Specifically, within states $s_{2-5}$ (i.e., $(0,1)$, $(0,2)$, $(0,3)$, $(0,4)$), where only predators are around, individuals exhibit a strong preference in the action of $\lambda=3$, the strongest migration willingness to escape. By contrast, when there are only prey around, i.e., the state of $(1,0)$, $(2,0)$, $(3,0)$, $(4,0)$, individuals are prone to stay put with the action of $\lambda=-3$, the weakest migration willingness. 
When the neighborhood is mixed with prey and predators, the two tendencies become compromised. For example, the escape willingness is weakened when there are also prey in their neighborhood, see the distributions within states $(1,1)$, $(1,2)$, and $(1,3)$.
In a special scenario of $s_1=(0,0)$, where neither prey nor predator is around, individuals prefer to move fast away, as no prey to feed themselves. These observations are consistent with the facts seen in nature.

The two tendencies are further strengthened when computing the probability density function (PDF) of different actions adopted in the evolution, aggregated across all states, see Fig.~\ref{fig:balanced}(b).
The PDFs of the three species are nearly identical, again confirming the preference symmetry among them. A key characteristic of these distributions is their bimodal profile: individuals prefer either fast or slow migration, while the intermediate migration actions, such as $\lambda=0, \pm1$, are less frequently adopted.
 It is this bimodal distribution that suppresses the formation of spiral waves, because when individuals are of different migration willingness, the foundation of clusters is ruined as they migrate differently. And this is the reason behind the unstructured patterns seen in Fig.~\ref{fig:pattern}(e-h).

Since the coexistence of ``survival-priority" and ``predation-priority" suppresses the spiral waves and thus sustains the biodiversity, the imbalance between the two tendencies is expected to jeopardize the stability of the ecosystem, as demonstrated below.

\begin{figure*}
\centering
\includegraphics[width=0.4\linewidth]{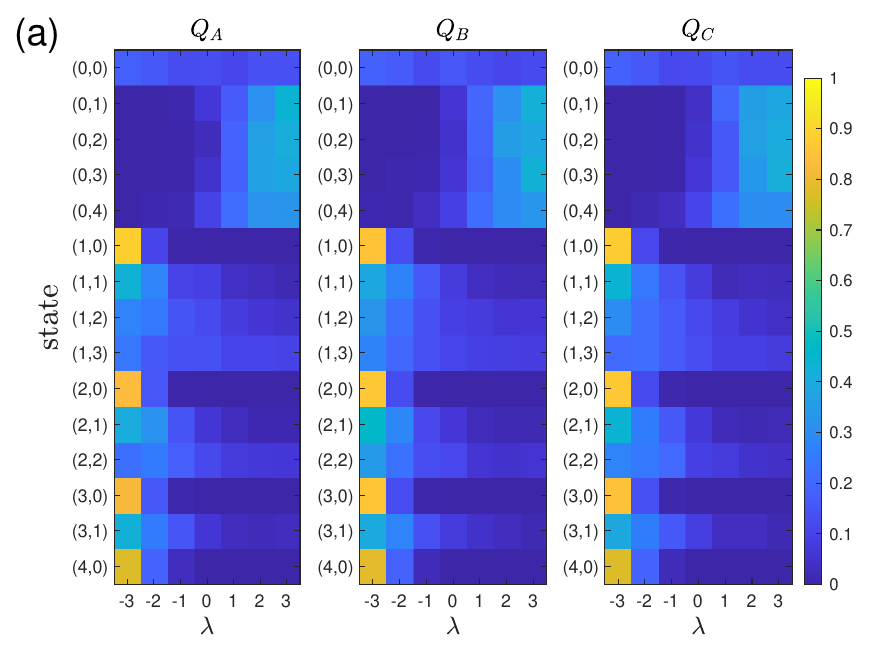}
\includegraphics[width=0.4\linewidth]{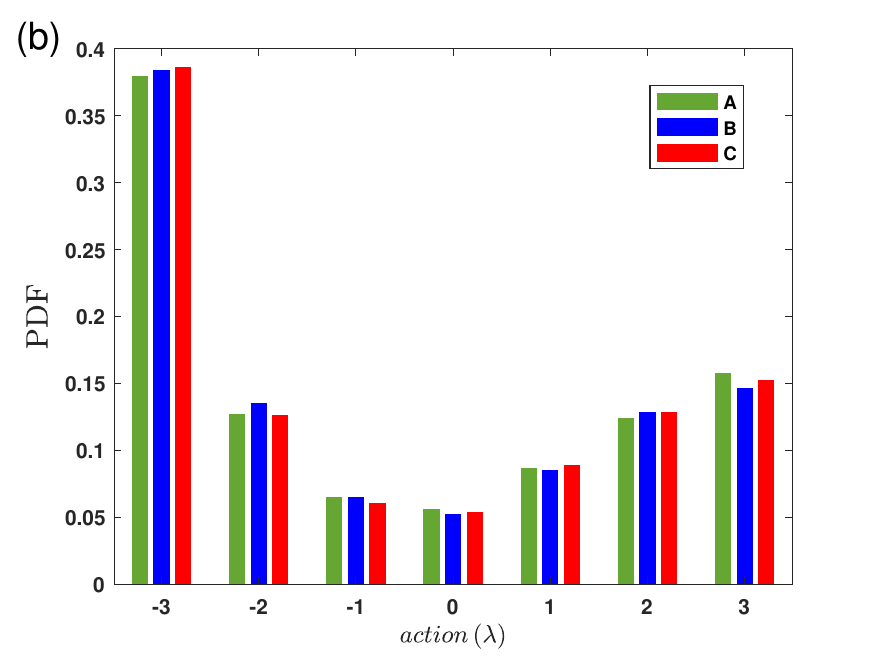}
\includegraphics[width=0.4\linewidth]{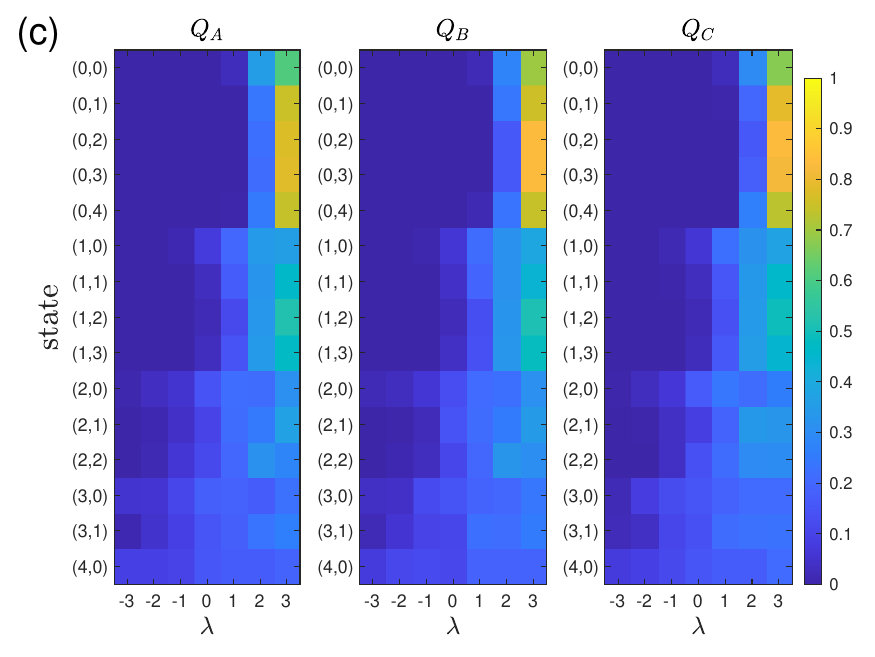}
\includegraphics[width=0.4\linewidth]{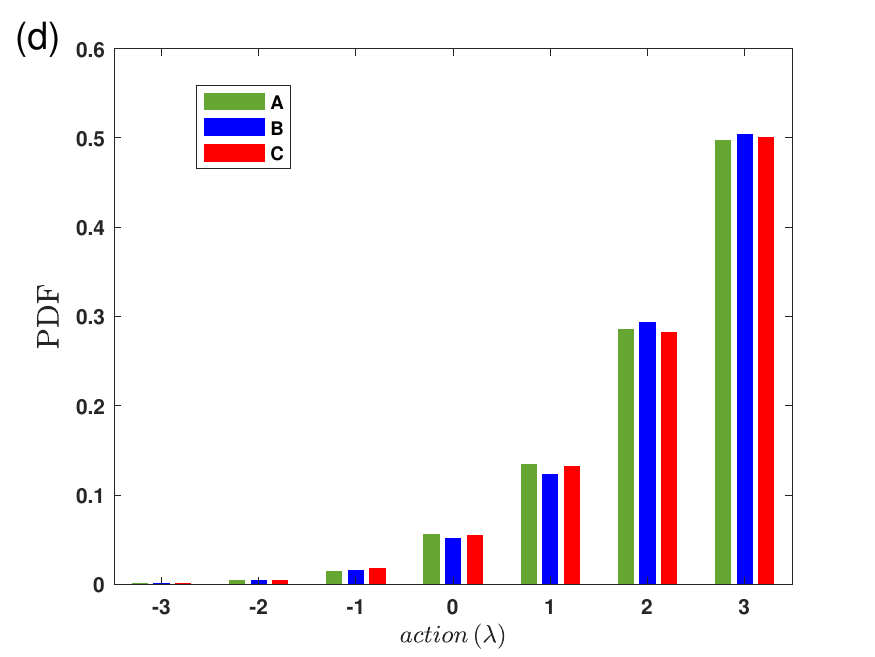}
\caption{\textbf{Distribution of action preferences in two imbalanced scenarios.} 
The top row (a, b) represents the predation dominance scenario ($R_{p}=40$ and $R_s=0.5$), and the bottom row (c,d) is for the survival dominance scenario ($R_{p}=1$ and $R_s=40$).
(a,c) Color-coded action preferences for the three species, {probabilities are defined in the same way as in  Fig.~\ref{fig:balanced}(a)}. 
(b,d) PDF of different actions adopted $\lambda$ for the three species in the stable state. 
Compared to the balanced scenarios shown in Fig.~\ref{fig:balanced}, the action preferences and thus the two tendencies are changed: while low mobility is enhanced in the predation dominance, the mobility is strengthened in the survival dominance is prioritized.
The settings are the same as Fig.~\ref{fig:balanced} except for the two rewards.
}
\label{fig:imbalanced}
\end{figure*}

\emph{Predation dominance --} When individuals over-emphasize predation, individuals put themselves at the risk of being predated as they may fail to escape in time.  Fig.~\ref{fig:imbalanced}(a,b) reports the distribution of action preferences in this scenario by raising the reward of a successful predation up to $R_{p}=40$. As can be seen, the preference in action $\lambda=-3$ is substantially strengthened in states where prey are present, such as $(1,0)$, $(2,0)$, $(3,0)$, and $(4,0)$, among others. In contrast, escape willingness is markedly reduced compared to the balanced case [Fig.~\ref{fig:balanced}(a)], particularly in states like $(0,1)$, $(0,2)$, $(0,3)$, and $(0,4)$. As shown in Fig.~\ref{fig:imbalanced}(b), individuals under predation dominance tend to remain stationary rather than migrate rapidly — a clear deviation from the distribution in the balanced scenario [Fig.~\ref{fig:balanced}(b)].

\emph{Survival dominance --} When survival is prioritized over predation, escape from predators becomes the top priority. Fig.~\ref{fig:imbalanced}(c,d) reports the results for this scenario with $R_{s} = 40$ and $R_{p} = 1$. As predation becomes less rewarding, the action of $\lambda=-3$ is rarely chosen, indicating that individuals prefer not to stay put to catch prey. Instead, actions with $\lambda>0$ (i.e., faster migration) become prevalent in almost all states, especially in the absence of prey (states $s_{1-5}$: $(0,0)$ to $(0,4)$).  The resulting action distribution becomes a single-peaked profile, and the average population mobility exceeds that of the above two scenarios (see Fig.~\ref{fig:imbalanced}(d)).

\begin{figure}[htpb]
\centering
\includegraphics[width=0.9\linewidth]{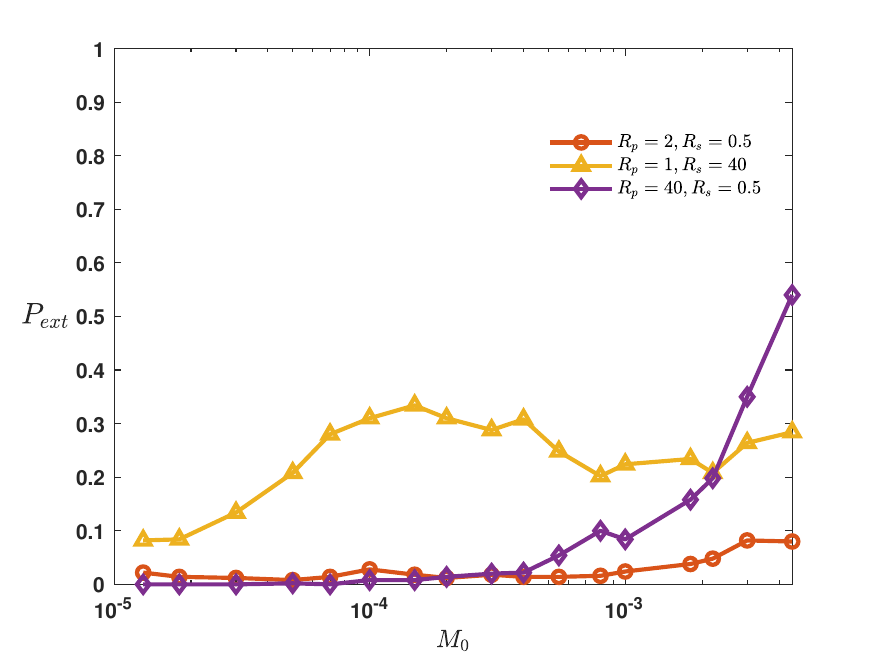}
\caption{\textbf{Extinction probability versus $M_{0}$ for the three scenarios:} 
the balanced scenario ($R_{p}=2$, $R_{s}=0.5$), the predation dominance ($R_{p}=40$, $R_{s}=0.5$), and the survival dominance ($R_{p}=1$, $R_{s}=40$). As seen, the extinction in both imbalanced scenarios is likely to occur.
Parameter: $N=100\times100$.
}
\label{fig:7}
\end{figure}

Figure~\ref{fig:7} further reveals that the extinction probability $P_{ext}$ rises once the balance between these two tendencies is disrupted.
In the predation dominance scenario, as many individuals prefer low migration, they are likely to aggregate and clusters are formed, leading to dynamics similar to those in spiral wave regimes (Appendix~\ref{Appendix:predation}) and consequently higher extinction likelihood. 
In the survival dominance scenario, the action distribution becomes unimodal, where evolution is then also reduced to the traditional scenario with high mobility, and extinction is thus expected. Here, the species with relatively smaller mobilities are put in a vulnerable position, triggering the extinction event. 
This underscores that migration decisions driven by a single objective — whether predation or survival — are insufficient to maintain biodiversity. A balance between both incentives is essential for individuals to adapt effectively to their environment and ensure species coexistence.

\begin{figure}[htpb]
\centering
\includegraphics[width=1.0\linewidth]{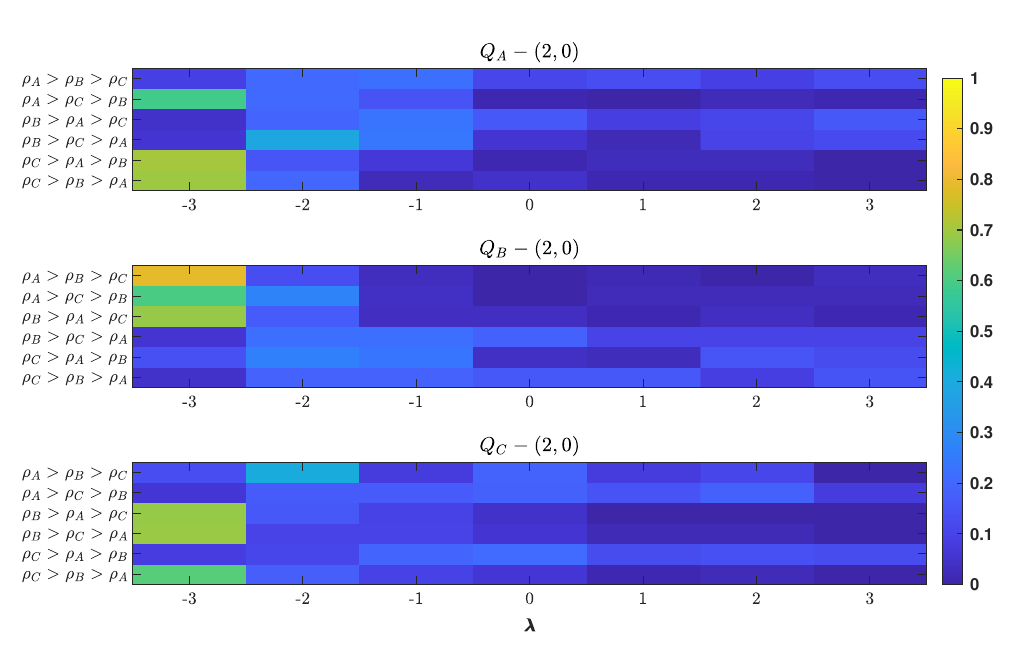}
\caption{\textbf{Symmetry-breaking in preferences within state $s_{10}=(2,0)$.} 
{Color-coded action preferences in the balanced scenario ($R_{p}=2$, $R_{s}=0.5$) for the species A, B, and C, respectively. Unlike Fig.~\ref{fig:balanced}(a), only the state $s_{10}=(2,0)$ is shown here, which is further divided into six subcategories based on the rankings in evolutionary outcome.}
The three Q-tables are not the same.
Parameters: $N=300\times300$ and $M_{0}=5\times 10^{-4}$.
}
\label{fig:state20}
\end{figure}
\section{Symmetry-breaking in densities}\label{sec:symmetry-breaking}
As shown in Fig.~\ref{fig:Pext}(b), the densities of the three species are not the same, which is surprising because their parameters are identical and their action preferences shown in Fig.~\ref{fig:balanced} are very similar. In Fig.~\ref{fig:Pext}(b), the three densities follow the order $\rho_A>\rho_C>\rho_B$, while all other five ranking orders are also equally likely to be observed in our stochastic simulations {(see Appendix~\ref{Appendix:six_ordering})}. How does the symmetry-breaking in densities occur?

Detailed examination reveals that the action preference patterns as shown in Fig.~\ref{fig:Pext}(b) are nearly identical except for the state $s_{10}=(2,0)$. A symmetry-breaking in action preference is revealed when we distinguish all six rankings and plot the preferences for each of the three species.
Fig.~\ref{fig:state20} displays their action preferences for these six subcategories, where the preference patterns are no longer identical. 
There are two qualitatively different classes: i) two species prefer $\lambda=-3$ (with orderings $\rho_A>\rho_C>\rho_B$, $\rho_B>\rho_A>\rho_C$, $\rho_C>\rho_B>\rho_A$), and ii) only one species does so for the rest three rankings. 
For the first class, considering $\rho_A>\rho_C>\rho_B$ (the one seen in Fig.~\ref{fig:Pext}(c)), it shows that for state $s_{10}$ the most preferred action for species A and B are both $\lambda=-3$, but this is not true for species C. When species A prefers $\lambda=-3$ within $s_{10}$, species C benefits the most because they are more likely to catch species A, and their predator (i.e., species B) suffers as they are predated by A, leading to $\rho_C>\rho_B$. Similarly, when species B prefers $\lambda=-3$, species A benefits most, and C suffers, resulting in $\rho_A>\rho_C$. This combined effect explains the density ranking $\rho_A > \rho_C > \rho_B$. 
An example from the second class is the subcategory with ranking $\rho_A>\rho_B>\rho_C$, where only species B prefers $\lambda=-3$. This preference benefits A the most and causes species C to suffer, leading to $\rho_A>\rho_B>\rho_C$ as a result.

\begin{figure}[tpb]
\centering
\includegraphics[width=0.85\linewidth]{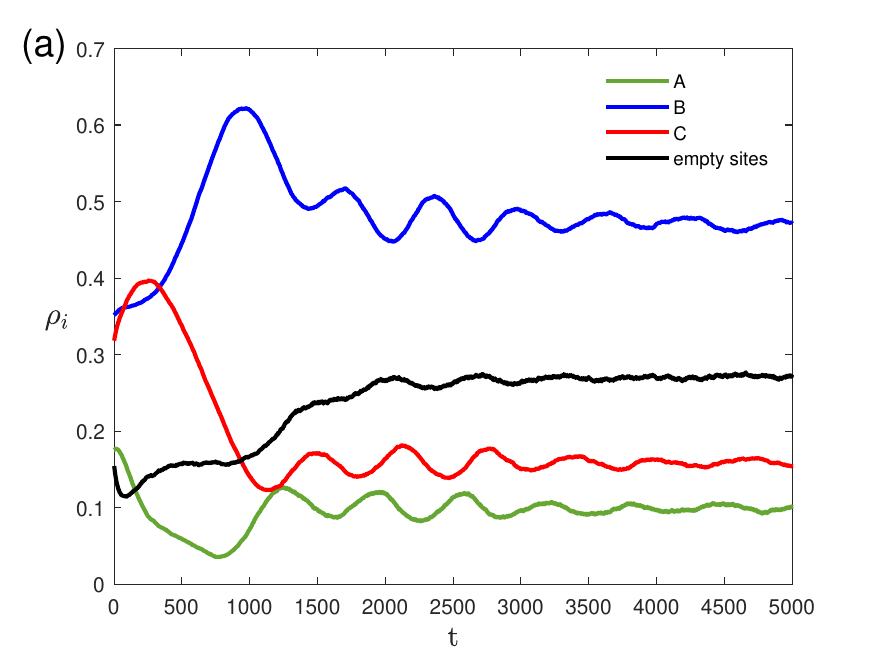} 
\includegraphics[width=0.85\linewidth]{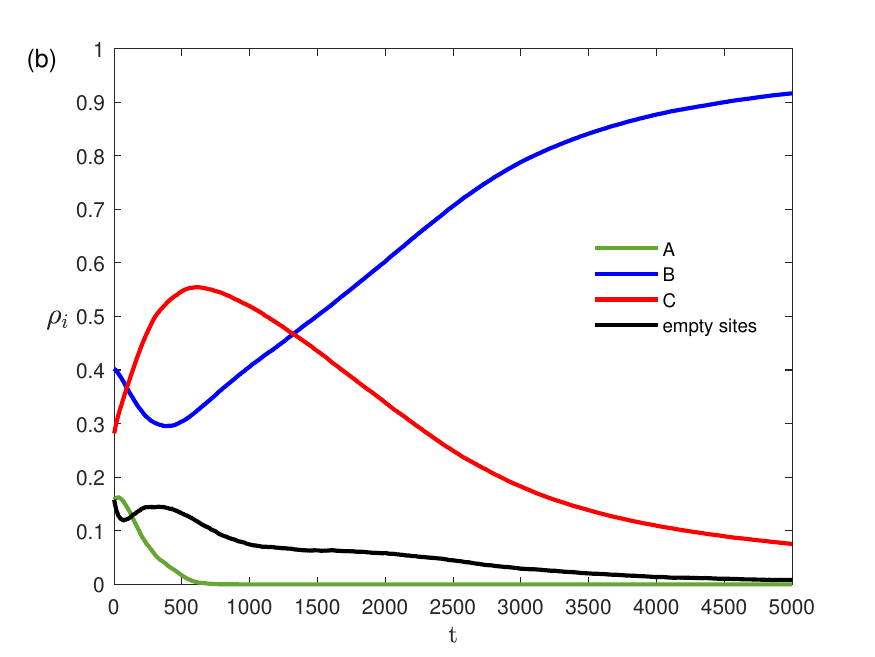}
\caption{\textbf{Temporal evolution of species densities for heterogeneous Q-learning species.} 
In such a mixture, species A, B, and C are balanced preference ($R_p=2, R_s=0.5$), survival dominance ($R_p=1, R_s=40$), and predation dominance ($R_p=40, R_s=0.5$), respectively. The three species could either coexist (a) or go extinct (b).
{By 1000 ensemble simulations, $31.5\%$ of runs evolve into the species coexistence, and the rest go extinct.}
Parameters: $N=300\times300$ and $M_{0}=5\times10^{-4}$.
}
\label{fig:ts3}
\end{figure}

\section{Extension}\label{sec:extension}

\subsection{Heterogeneous Q-learning species}
While the above investigation assumes a uniform parameterization for all three species, different species in the real world generally have different parameters. As an example, here we examine a typical asymmetrical scenario where species A has a balanced preference ($R_{p}=2, R_{s}=0.5$), B is survival-dominant ($R_{p}=1, R_{s}=40$), and C is predation-dominant ($R_{p}=40, R_{s}=0.5$). We are interested in how these reward differences impact the evolution of the population and biodiversity.

Simulations show that, in such an asymmetric scenario, the extinction probability is significantly higher (e.g., rising to $67.8\%$ for the given parameters in Fig.~\ref{fig:ts3}) than the near-zero probability in the symmetrical setup shown in Fig.~\ref{fig:Pext}.
Two typical time series of species densities are presented, respectively, for coexistence and extinction, in Fig.~\ref{fig:ts3}(a) and~\ref{fig:ts3}(b). In both cases, species B clearly holds a density advantage. This is because, on the one hand, survival dominance makes them more likely to escape from predators A; on the other hand, predation dominance of species C makes them easier for species B to catch.
Unexpectedly, species A with a balanced preference performs the worst. This is reasonable as their prey (species B) are survival-dominant, making them hard to catch due to their strong escape tendency, while their predators (species C) are predation-dominant, making them difficult to escape. This renders species A the most vulnerable, which could lead to the collapse of the ecosystem, as shown in Fig.~\ref{fig:ts3}(b). 

\subsection{Q-learning versus traditional species}
\begin{figure}[tpb]
\centering
\includegraphics[width=0.85\linewidth]{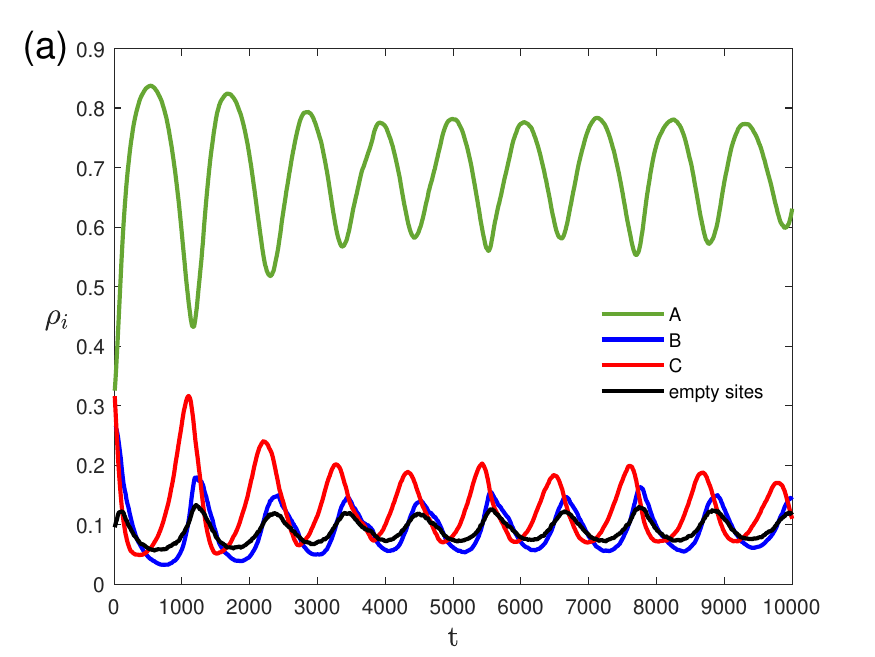} 
\includegraphics[width=0.85\linewidth]{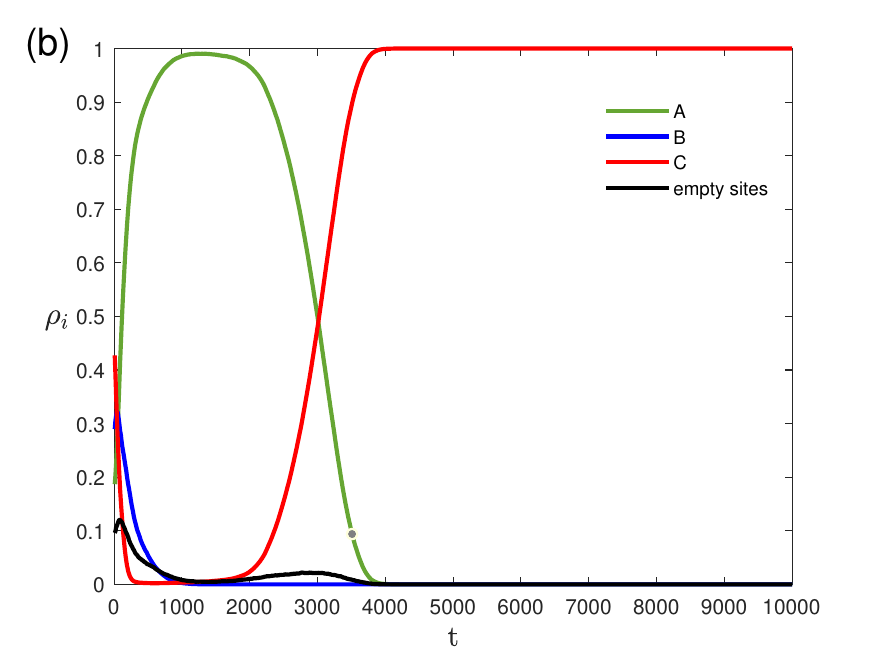}
\caption{\textbf{Temporal evolution of species densities for a mixed scenario.} 
In such a mixture, A is empowered by Q-learning ($R_p=2, R_s=0.5$), B and C are traditional species with constant migration rates ($\lambda=0$).
{By 1000 ensemble simulations, $80.9\%$ of the evolve into the species coexistence, and the rest go extinct.}
Two typical density evolutions are shown for coexistence (a) and extinction (b), respectively.
Parameters: $N=300\times300$ and $M_{0}=3\times10^{-5}$.
}
\label{fig:QvsRMF}
\end{figure}

Another interesting setup involves combining traditional species with Q-learning type to see whether adaptivity offers any advantage over species with constant migration. Fig.~\ref{fig:QvsRMF} displays the evolutionary outcome for a scenario where species A is of Q-learning type, while B and C have constant migration rates (i.e., $\lambda=0$).

Simulations show that the extinction probability also increases ($14.8\%$ for the given parameter in Fig.~\ref{fig:QvsRMF}) in such a mixed scenario compared to the pure Q-learning setup shown in Fig.~\ref{fig:Pext}. {As seen in the typical time series, species A, empowered by Q‑learning, demonstrates a significant advantage over the two traditional species (B and C). This advantage is so pronounced that species A can be driven to extinction by its predator (species C), ultimately leading to the collapse of the entire system — see Fig.~\ref{fig:QvsRMF}(b).} Otherwise, coexistence is maintained with species A dominating, but with strong oscillations. Interestingly, in the extinction case shown in Fig.~\ref{fig:QvsRMF}(b), after species B disappears, A's predator (species C) rises to a high level, and ultimately, the Q-learning species A also vanishes.

\section{Conclusion}\label{sec:conclusion}

Driven by the mystery of biodiversity, particularly how highly mobile species coexist within a spatial domain, we propose a reinforcement learning paradigm to understand how different species coexist even when they migrate rapidly. Specifically, we investigate a population of three species with the RPS cycling dominance structure, where a joint Q-learning algorithm adaptively adjusts their mobility. Each species is associated with a Q-table that guides the movement of its individuals. This model emphasizes the adaptability of migration, allowing the adjustment of mobility based on the environment, rather than being fixed as most previous studies assumed~\cite{2007Mobility}. We show that the adaptation enabled by joint Q-learning allows all three species to coexist effectively, as the extinction probability remains quite low across a wide range of baseline migration rates. Their action preference patterns reveal that they develop two main tendencies: survival-priority — escaping from predators, and predation-priority — staying put to catch prey, both of which are commonly observed in nature. Mechanistically, the two tendencies create heterogeneities in mobility, disrupting the formation of spiral waves and thus promoting species coexistence. However, the imbalance of the two tendencies reduces mobility heterogeneity and therefore jeopardizes the ecosystem’s biodiversity.

Notably, we observe a symmetry-breaking in action preference within state (2,0), where the three species develop distinct behavioral patterns. This subtle differentiation ultimately leads to divergent species densities. Further investigations show that when conventional fixed-migration species are mixed with Q-learning agents, the latter gain an evolutionary advantage due to their adaptability. Mixing also introduces heterogeneity among Q-learning species, where a rich spectrum of dynamical phenomena is observed.

{Biologically, the two migration tendencies resemble the heuristic rule proposed in Ref.~\cite{2013Persistent}, but here they are learned via Q‑learning rather than specified a priori. Dynamically, the mobility heterogeneities in Ref.~\cite{2013Persistent} are not strong enough to suppress spiral wave formation, which sustains a higher likelihood of extinction.}

Methodologically, our joint Q-learning framework developed here aligns with previous studies~\cite{Ding2023Emergence, zheng2024evolution, Zheng2024decoding, zheng2025decoding, zheng2025optimal} in its core conception, but differs in two key aspects. First, learning operates at the species level rather than the individual level, reflecting the ecological reality that adaptive behaviors in many species arise collectively and are inherited across generations. 
 {Note that learning at the species level remains coarse-grained, as it does not account for spatial localization. A natural extension would be a multi-patch description in which individuals within a patch share a common Q-table, while Q-tables differ across patches, with spatial coupling between patches -- similar to meta-population models used in epidemic spreading studies~\cite{Murray2003mathematical}.} 
 Second, the Q-table is first trained and then applied to guide actions without further updates — a departure from the typical co-evolution of learning and gaming processes. This simplification is motivated by time-scale separation: learning often occurs over evolutionary timescales, resulting in relatively stable collective strategies once the system approaches equilibrium. {As an extension, it may also be interesting to develop the co-evolutionary framework in the future, where the two dynamical processes unfold simultaneously.}

In summary, given the prevalence of learning in living organisms, reinforcement learning offers a more natural paradigm for studying ecological evolution than traditional mechanistic models with fixed rules and parameters. However, to establish RL as a robust framework for understanding biodiversity and ecosystem evolution, empirical field experiments are essential to validate its underlying logic~\cite{Underwood1997experiments}.

\section*{Data and code availability}
All data in our study is generated by our codes written in Matlab, which are available at \href{https://github.com/chenli-lab/RL-RPS}{https://github.com/chenli-lab/RL-RPS}.

\section*{ACKNOWLEDGEMENTS}
This work is supported by the National Natural Science Foundation of China (Grants Nos. 12075144, 12165014), the Fundamental Research Funds for the Central Universities (Grant No. GK202401002), and the Key Research and Development Program of Ningxia Province in China (Grant No. 2021BEB04032). We acknowledge enlightening discussions with Uwe C. T\"{a}uber (Virginia Tech) during his visit to SNNU.

\appendix

\section{The protocol of Q-learning and pseudocode}\label{Appendix:pseudocode}

To sum up, the protocol of our joint Q-learning version of the spatial RPS model can be summarized as follows:
\begin{itemize}
\item [1)]
Each site of the lattice is randomly occupied by an individual of A, B, C, or left empty with equal chance.
 Initialize all the items of the three Q-tables with random numbers $Q_{s,a}\in(0,1)$ independently to mimic the unawareness of individuals to the surroundings. Each individual $i$ takes a random migration rate with $a_i\in\mathcal{A}$. 
\item [2)] In the learning process, each agent's action is made by pure exploration $a_i\in\mathcal{A}$; afterwards, their rewards are obtained by collecting payoffs, and then they update their Q-tables to accumulate their experience. Their states also need to be updated. 
\item [3)] After the three Q-tables are converged, the gaming process starts. Their migration is then strictly guided by the corresponding Q-table belonging to their species, and the three Q-tables are no longer revised. 
 \end{itemize}
 Repeat step 2 till the convergence of the three Q-tables, which completes the learning process. Repeat step 3 until the system reaches a statistically stable state or the desired time duration.
The pseudocode is provided in Algorithm~\ref{algorithm:1}, which offers more simulation details.

\begin{algorithm}
	\SetAlgoNlRelativeSize{0} 
	\SetInd{0.5em}{0.5em} 
	\SetAlgoNlRelativeSize{-1} 
	\SetNlSty{text}{}{\hspace{0.5em}} 
	\caption{RPS model with joint Q-learning}
	\KwIn{$\alpha,\gamma$}
	Initialization\;
    \quad $Q_{1}, Q_{2}, Q_{3} \gets \text{random}(15 \times 7)$\; 
    \quad $\text{Lattice sites} \gets \text{random}[0,3]^{L \times L}$\; 
    \quad $\sigma,\mu \gets \text{1}$\; 
    \quad $N_{step} \gets \text{10}$; 
    
    Learning Process\;
	
	\Repeat{the termination condition is met}{
		\For{each round $t$}{
		
			\For{Each agent}{								
					Agent picks a random action $a\in\mathbb{A}$\;				
				}

			\For {interaction count $= 1$ to $N_{step} \times L^2$}{
				Randomly select an agent and its neighbor\;
				$r=rand()$\;
				\If{$r< \sigma /(\sigma + \mu+ \varepsilon)$}{
				      Reactions~(\ref{eq:1}) if available\;
				}
				\ElseIf{$\sigma/(\sigma + \mu+ \varepsilon)<r< (\sigma + \mu)/(\sigma + \mu+ \varepsilon)$}{
				      Reactions~(\ref{eq:2}) if available\;
				}
				\Else{
				     Reactions~(\ref{eq:3})\;
			    }
			}
			
			\For{Each agent}{
				Calculate the reward $R$ for each agent\;
				Update $s$\;
				Update $Q_{1}, Q_{2}, Q_{3}$ according to Eq.~(\ref{eq:Bellman_eq})\;
		    }
		}
	}
	
	Gaming Process\;
	\Repeat{the termination condition is met}{
		\For{each round $t$}{
			
			\For{Each agent}{								
				Agent acts according to the Q-table of its species\;				
			}	
			
			\For {interaction count $= 1$ to $N_{step} \times L^2$}{
				Randomly select an agent and its neighbor\;
				$r=rand()$\;
				\If{$r< \sigma /(\sigma + \mu+ \varepsilon)$}{
				      Reactions~(\ref{eq:1}) if available\;
				}
				\ElseIf{$\sigma/(\sigma + \mu+ \varepsilon)<r< (\sigma + \mu)/(\sigma + \mu+ \varepsilon)$}{
				      Reactions~(\ref{eq:2}) if available\;
				}
				\Else{
				     Reactions~(\ref{eq:3})\;
				}
			}			
		}
	}	
\label{algorithm:1}
\end{algorithm}

In the learning process (i.e. step 2), the two learning parameters are fixed at $\alpha=0.1$ and $\gamma=0.9$, a typical parameter combination~\cite{Ding2023Emergence, zheng2024evolution} where the species both appreciate historical experience and hold long-term vision in decision-making.

\section{Evolution in the scenario of predation dominance}\label{Appendix:predation}

\begin{figure}[tb]
\centering
\includegraphics[width=0.8\linewidth]{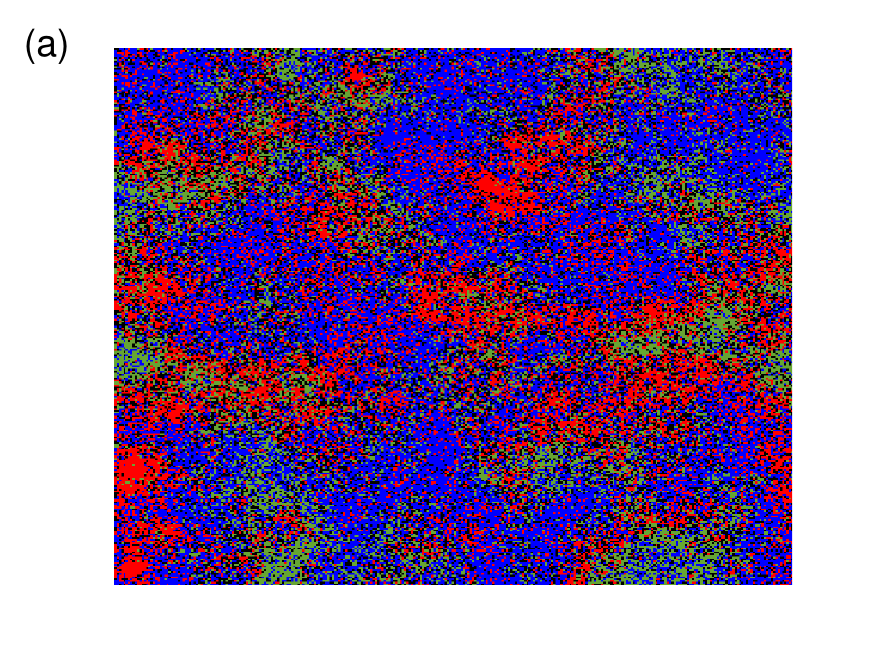} 
\includegraphics[width=0.8\linewidth]{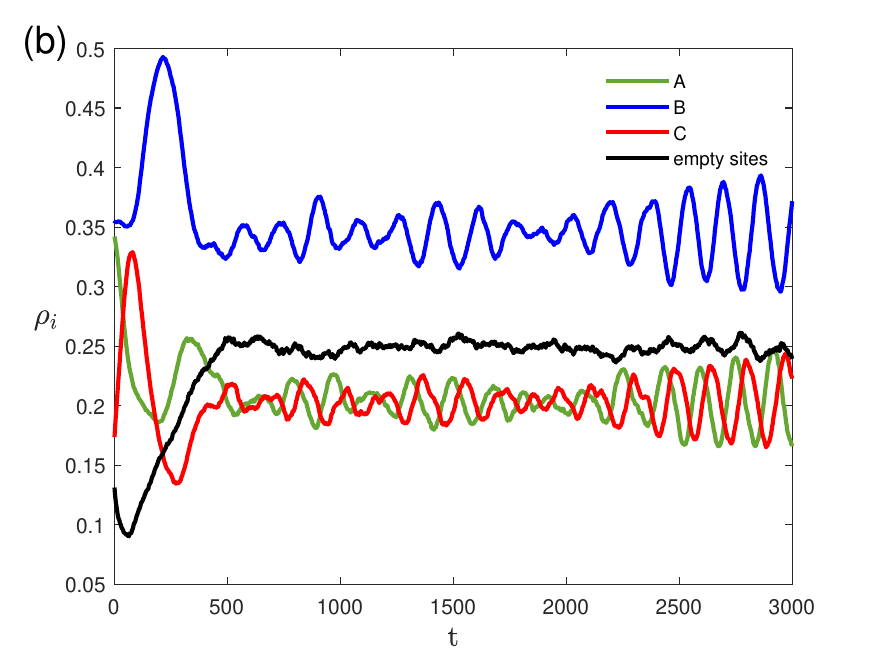}
\caption{\textbf{Evolution in predation dominance scenarios.} 
The typical pattern (a) and time series (b) in the predation dominance scenario with $R_p=40$ and $R_s=0.5$.
Parameters: $N=300\times300$ and $M_{0}=5\times10^{-4}$.
}
\label{fig:predation}
\end{figure}

To better understand the evolution in predation dominance scenarios, we provide the typical pattern and time series for parameters $R_p=40$ and $R_s=0.5$.
Fig.~\ref{fig:predation}(a) shows that, due to the tendency to stay put,  individuals aggregate and form some clusters, though not as compact as the patterns observed in the RMF model (e.g. Fig.~\ref{fig:pattern}(a-c)). Time series shows that after the transient, oscillation emerges in the densities of three species -- a signature for the spiral wave (e.g Fig.~\ref{fig:Pext}(b)). 

\section{{Probabilities of six orderings}}\label{Appendix:six_ordering}

\begin{figure}[tpbh]
\centering
\includegraphics[width=0.9\linewidth]{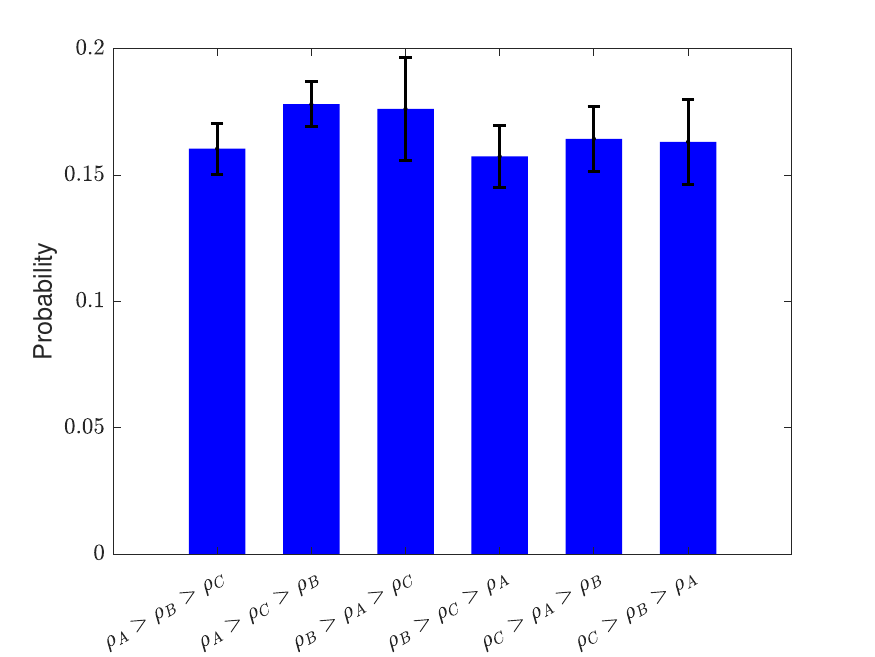} 
\caption{
{\textbf{Probabilities of six orderings.} 
The probabilities within state $s_{10}=(2,0)$ with 6000 independent realizations are conducted, with the error bars being shown for each ordering.
Parameters: $R_{p}=2$, $R_{s}=0.5$, $M_{0}=5\times10^{-4}$, and $N=300\times300$.}
}
\label{fig:six_orderings}
\end{figure}

{As shown in Sec.~\ref{sec:symmetry-breaking}, six distinct orderings emerge in the simulations through symmetry-breaking. To further characterize how frequently these orderings occur, we conduct 6000 independent simulations with the same parameters as in Fig.~\ref{fig:state20}. As shown in Fig.~\ref{fig:six_orderings}, each of the six orderings appears with approximately equal probability, indicating that the symmetry-breaking process does not favor any particular outcome.}

\bibliography{reference}

\end{document}